\documentclass[
aps,%
12pt,%
final,%
notitlepage,%
oneside,%
onecolumn,%
nobibnotes,%
nofootinbib,%
superscriptaddress,%
noshowpacs,%
centertags]%
{revtex4}

\begin{document}


\title
{PRINCIPAL COMPONENT ANALYSIS FOR ESTIMATING PARAMETERS
OF THE L1287 DENSE CORE BY FITTING MODEL SPECTRAL MAPS INTO OBSERVED ONES}

\author{\firstname{L.~E.}~\surname{Pirogov}},
\email{pirogov@appl.sci-nnov.ru}
\affiliation{%
Institute of Applied Physics, Russian Academy of Sciences, Nizhny Novgorod, Russia}
\author{\firstname{P.~M.}~\surname{Zemlyanukha}},
\affiliation{%
Institute of Applied Physics, Russian Academy of Sciences, Nizhny Novgorod, Russia}

\begin{abstract}

An algorithm has been developed for finding the global minimum of a multidimensional
error function by fitting model spectral maps into observed ones.
Principal component analysis is applied to reduce the dimensionality of the model
and the coupling degree between the parameters, and to determine the
region of the minimum.
The $k$--nearest neighbors method is used to calculate the optimal parameter values.
The algorithm is used to estimate the physical parameters of the contracting
dense star-forming core of L1287.
Maps in the HCO$^+$(1--0), H$^{13}$CO$^+$(1--0), HCN(1--0), and H$^{13}$CN(1--0) lines, calculated
within a 1D microturbulent model, are fitted into the observed ones.
Estimates are obtained for the physical parameters of the core, including
the radial profiles of density ($\propto r^{-1.7}$), turbulent velocity ($\propto r^{-0.4}$),
and contraction velocity ($\propto r^{-0.1}$).
Confidence intervals are calculated for the parameter values.
The power-law index of the
contraction-velocity radial profile, considering the determination error,
is lower in absolute terms than the
expected one in the case of gas collapse onto the protostar in free fall.
This result can serve as an argument in favor of a global contraction model
for the L1287 core.

\end{abstract}

\maketitle

\newpage

\section{Introduction}

Studies on the structure and kinematics of the dense cores of molecular clouds
provide information on the initial conditions and early stages 
of the star-formation process to be utilized in theoretical models.
This is especially important when studying the regions of massive star and
star cluster formation, which evolutionary scenarios are now only beginning
to develop (see, e.g., [1, 2]).

According to observational data, massive stars ($\ga 8$~M$_{\odot}$) and
star clusters are formed in near-virial equilibrium dense cores that are located
in filament-shaped massive gas-dust complexes and clumps (see, e.g., [3--7]).
The existing theoretical models of star formation
employ different assumptions about the initial core state.
Thus, the isothermal sphere model, which is applied for describing
the formation of isolated low-mass stars [8, 9], assumes that the quasi-equilibrium
spherical core with a Bonnor-Ebert-type density profile (a flat segment near
the center and a near-$\propto r^{-2}$ dependence in the envelope) evolves
towards a singularity at the center (protostar), after which a collapse begins,
which propagates ``inside-out''.
The turbulent-core model [10], proposed for describing the formation of massive
stars and star clusters, also considers, as the initial state,
a hydrostatic-equilibrium sphere characterized by supersonic turbulence and
a $\propto r^{-3/2}$ density profile [10, 11].
Both the isothermal sphere model and the turbulent core model use density
and velocity profiles in the region where gas collapses onto the star
of the form $\propto r^{-3/2}$ and $\propto r^{-1/2}$, respectively.
As shown in [12], these profiles do not depend on the state of gas in the core.

An alternative model of global hierarchical collapse [13] proceeds from
the fact that the cores, like the parent clouds, are nonequilibrium objects
that are in an ongoing process of global collapse even before the protostar
formation, and their observed closeness to virial equilibrium is due,
specifically, to the closeness of the free fall velocity to the virial one.
In this model, which is based on the classical works of Larson and Penston
[14, 15], after the formation of the protostar, the density profile in the
envelope becomes $\propto r^{-2}$ and the contraction velocity is independent
of the radial distance (see, e.g., [16, 13]). Near the protostar, where
the collapse occurs, the radial profiles of density and contraction velocity
are the same as in the isothermal-sphere and turbulent-core models.
Thus, the information about the density profile is insufficient for us to make
a choice between the above models; firstly, we need to know the velocity profile
in the outer regions of the cores.

The kinematics of the cores is estimated mainly from observations
in molecular lines. The presence of systematic velocities along the line
of sight leads to a shift in the centers of optically thick and thin lines
(see, e.g., [17]) and to the appearance of asymmetry in the spectra of optically
thick lines due to the absorption of the emission from the inner layers
by outer ones and due to the Doppler effect (see, e.g., [18, 19]).
The average contraction velocity of the core can be estimated
within more or less simple models from asymmetric line observations at one point
(see, e.g., [20--23]). To estimate the radial profile of systematic velocity,
it is necessary to fit the model spectral maps into the observed ones,
while simultaneously calculating or setting the profiles of the other physical
parameters.

Automatic fitting methods of model spectral maps into observed ones
in the case of several free parameters are rarely used today. Researchers
usually compare the spectra observed at individual points with simulated ones;
less often, they use spectral maps, varying one or two parameters and considering
the remaining ones to be specified from independent observations, theoretical
model predictions, or preliminary calculation results (see, e.g., [24--29]).
In this case, researchers either consider the systematic contraction velocity
to be constant or use a radial profile $\propto r^{-1/2}$.
Finding the optimal values while varying several parameters simultaneously
may be difficult because the error function may have more than one local
minimum and the parameters themselves may correlate with one another, leading
to dependence on the initial conditions and to poor convergence.
The use of special methods to search for the global minimum of the error
function (e.g., the method of differential evolution [30]) in the case of
a model with several free parameters may lead to considerable computational
costs.

In recent years, $Principal~Component~Analysis$ (PCA) has been successfully
applied to studying experimental data [31].
Within this method, data are transformed to such an optimal basis in which
linear relations between the basic vectors are excluded. This approach allows
one to reduce the dimensionality of the data. This method is quite often used
to reduce the dimensionality of astronomical data (see, e.g., [32] and references
therein), but it can also be applied to the results of model calculations
by reducing the dimensionality of the model and determining the range of parameter
values near the minimum of the error function. The exact values of the model
parameters, which correspond to the minimum of the error function, can be calculated
 by the regression method. For instance, the $k$--nearest neighbors ($k$NN) method
[33] appears to fit this purpose. It is an analogue of the least-squares
method, but unlike the latter, it allows one to choose, from a set of models,
only ones that correspond to observational data by the least-squared error
criterion.

This work aims to develop an algorithm that uses PCA and $k$NN to fit
model spectral maps into observed ones and to apply this algorithm for estimating
the radial profile of systematic velocity and other physical parameters
of the L1287 dense core. In this object, a cluster of low- and intermediate-mass
stars is being formed, and the observed profiles of optically thick lines
show an asymmetry pattern which indicates contraction (see, e.g., [34]).
The analysis used observational data in the lines of HCO$^+$(1--0) and HCN(1--0),
which are indicators of high-density gas ($\ga 10^5$~cm$^{-3}$ [18]) and
the isotope lines of these molecules. Observations in different lines
of the HCO$^+$ and HCN molecules are often used to search for massive cores
with systematic motions in the line of sight (see, e.g., [35--39]).
The HCN(1--0) line is, however, less often used for these purposes.
It has three hyperfine components with different optical depths and intensity
ratios that differ from the case of local thermodynamic equilibrium (LTE).
The observed profiles of these components may overlap.
To determine parameters of the physical and kinematic structure of the cores
from the HCN(1--0) data it is necessarily to use non-LTE models
(see, e.g., [40, 41]).
In this work, we calculated the excitation of HCO$^+$, HCN, and their isotopes
using a 1D microturbulent
spherically symmetric non-LTE model, the physical parameters of which,
including the systematic
velocity, were functions of the distance from the center.

	This paper consists of five sections and Appendix. Section 2 presents
the algorithm for fitting model spectral maps into observed ones using PCA and
$k$NN. Section 3.1 summarizes the observational data and physical properties
of the L1287 core. Section 3.2 describes the application of the algorithm
to the observational data on L1287 and the results of estimating the physical
parameters. Sections 4 and 5 present the results and discussion
and the conclusions, respectively.
A description of the model is given in Appendix.

\section{PCA-BASED ALGORITHM FOR FITTING MODEL SPECTRAL MAPS
INTO OBSERVED ONES}

The process of fitting model spectral maps into observed ones by means
of conventional iterative methods for estimating physical parameters
is complicated by the fact that the multidimensional error function (the total
discrepancy between the observed and model spectra) may have several local minima,
which creates a dependence on the initial values.
In this case, a correlation between the parameters may seriously worsen
the convergence.
Another approach is to calculate a set of model maps in advance for a grid
of model parameters and select those that are close to the observed parameters.
This is also a complicated approach because calculations for a discrete
$n$-dimensional grid (where $n$ is the number of parameters) that is
densely enough to cover the space of probable values may be beyond
the computational capabilities.
However, such a grid would obviously be redundant.
If we specify the parameter values randomly, then, with calculated model maps
for them, we can roughly determine a region within which lies the minimum
of the error function.
If we apply a certain transformation to the resulting region that minimizes
the relations between the parameters and transform it to a new space
of orthogonal vectors, we can reduce the dimensionality by discarding
the vectors carrying minimum information about the model parameters.
If we then fill the remaining vector space with a sufficiently dense grid and
make the inverse transformation, we obtain a filled-in space of model parameters
near the minimum, the exact value of which can be found by the regression method.
One such transformation can be PCA, a classical method of dimensionality
reduction [31]. It involves finding such a linear transformation where
the initial set of parameters is represented by a vector basis (with principal
components as the vectors), the correlations between the vectors being minimized.

The described general principles enabled the development of an algorithm
for finding the physical parameters of dense cores of molecular clouds
by fitting model maps of the molecular lines into the observed ones.
The algorithm involved a preliminary analysis of observational data and
determination of the to-be-free parameters, PCA-based dimensionality reduction
and determination of the region of model parameters near the minimum, and
finding the optimal values of free parameters by the $k$NN method [33] and
determination of the confidence region boundaries for each of them.
The diagram of the algorithm is shown in Fig. 1.
The optimal parameters were determined minimizing the error function:

\begin{equation}
\chi^2=\frac{1}{N_{p}-n}
\sum_{j=1}^{N} \sum_{i=1}^m
\frac {(I_{ij}^{obs}-I_{ij}^{mod})^2}{\sigma_{j}^2} \hspace{2mm},
\end{equation}
where $N$ is the number of spatial points in the map;
$m$ is the number of channels in the spectrum;
$I_{ij}^{obs}$ and $I_{ij}^{mod}$ are the observed and model intensities
in the $i$th spectral channel for the $j$th point in the map, respectively;
$\sigma_j$ is the standard deviation of the observed
spectrum at point $j$, calculated from intensity fluctuations
outside the line range; $N_{p}=m\times N$;
and $n$ is the number of parameters in the model.

\begin{figure}[t!]
\setcaptionmargin{5mm}
\onelinecaptionsfalse
	\includegraphics[width=\linewidth]{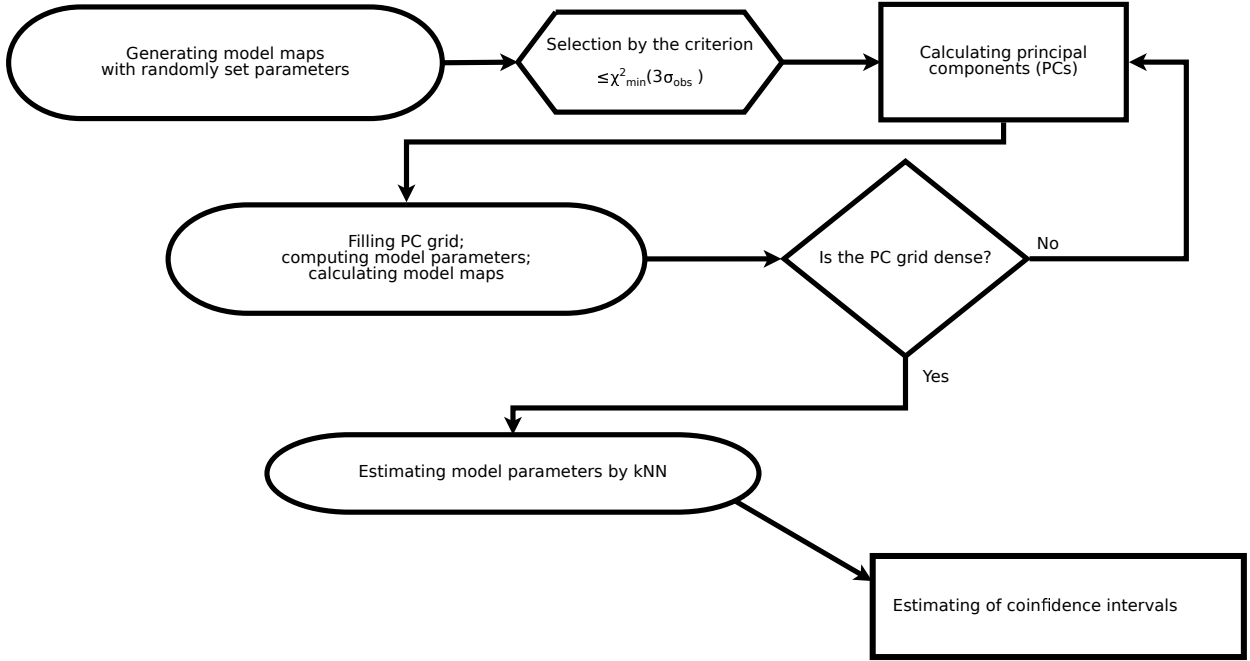}
\captionstyle{normal}
\caption{
Parameter determination diagram for fitting model spectral maps into
observed ones.}
\label{fig:diagram}
\end{figure}

In the course of preliminary analysis, we determined
the coordinates of the map's central point (the
center of the core) and the source velocity from the
observational data in the optically thin line.
Using a random number generator, we then created a set of
model parameters with sufficiently wide ranges and
calculated the model spectral maps and values of the error function.
Within this set, we selected those
parameters that satisfy the inequality $\chi^2\le \chi_{min}^2(3\sigma_{obs})$,
where $\chi_{min}^2(3\sigma_{obs})$ is the value of the error function at
those model parameters that yield the minimum value $\chi^2$
for a given set when noise with a standard deviation of $3\sigma_{obs}$ 
is added to the observed intensities.
For the resulting parameter sample, we calculated
matrices of the direct and inverse transformation into
the PC-space, reduced the number of components,
and filled the remaining space with a regular grid.
The grid nodes were transformed using the inverse transformation
into the values of the physical parameters.

Choosing a sufficient number of PCs is not a simple
problem since any dimensionality reduction
method causes information loss. An overview of possible
options for solving this problem is given in Appendix
to [42] and in the references to that paper.
In linear PCA, which we applied, the loss of information can
manifest itself in biased values of the physical parameters
after the direct and inverse transformations.
The number of remaining PCs, which determines the
extent of the loss, was chosen in such a way that the
ratio of the eigenvector sum for the PC covariance
matrix to the eigenvector sum for the covariance
matrix of the physical parameters differed from unity
(the value in the case of the identity transformation) by
no more than 10\% and the bias in the parameter estimates
did not create systematic errors [42, 43].

The final step was to calculate the physical parameter
values corresponding to the exact minimum of the
error function and estimate the errors. To this end, we
used the $k$NN method [33], which we previously
applied to estimate the physical parameters of the dense core of S255N [44].
The $k$NN method is similar to the least-squares method, but unlike the latter,
it does not adjust the model parameters to the
observed spectra but calculates the optimal parameter
values from the previously obtained spectra by the same criterion.
This method enables regression analysis
between a set of model maps with different parameters and the observed map.
Thus, among all the model maps, we found $k$ nearest ones to the observed
map by the criterion of the minimum of the mean $\chi^2$ value.
When there was no such minimum ($\chi^2$ increases with increasing $k$,
and averaging across the models increases the error
function), a denser grid was needed for PCs in the
region near the supposed minimum.
The optimal value of the $p$ parameter was the $\chi^2$-weighed mean
over $k$ selected instances:

\begin{equation}
p=\frac {\sum_{i=1}^{\rm k} p_i/\chi_i^2}{\sum_{i=1}^{\rm k} 1/\chi_i^2} \hspace{2mm},
\end{equation}
where $p_i$ and $\chi_i^2$ are the values of the parameter and
error function for the $i$th point in the parameter space,
respectively.

Using maps of the object in several spectral lines of
different optical depths, we can narrow down the
range of confidence values by fitting the model spectra
into these maps simultaneously. In this case, additional
model parameters are the relative abundances of the molecules.
The total error for the maps in several
lines ($n_{lines}$) is written as

\begin{equation}
\chi^2=\frac{1}{N_{p}-n}\sum_{k=1}^{n_{lines}}
\sum_{j=1}^{N_k} \sum_{i=1}^{m_k}
\frac {(I_{ijk}^{obs}-I_{ijk}^{mod})^2}{\sigma_{jk}^2} \hspace{2mm},
\label{allchi}
\end{equation}
where $N_{p}=\sum_{k=1}^{n_{lines}}N_k\times m_k$;
$N_k$ is the number of spatial points in the map in the $k$th line;
$m_k$ is the number of channels in the spectrum of the $k$th line;
and $\sigma_{jk}$ is the standard deviation of the observed spectrum in
the $k$th line at point $j$.

Since the parameter space is curvilinear, the confidence
regions for the probable parameter values were
determined by applying a cross-section of the multidimensional
error function by the hyperplane $\chi^2=\chi^2_{\sigma}$.
The calculation of $\chi^2_{\sigma}$ does not depend on the choice of the
basis; it is convenient to perform it in the PC space.
The threshold value was
$\chi^2_{\sigma}=\chi_{min}^2(pc_l^{opt}\pm\sigma_{pc_l})$,
i.e., the value of the error function in the case where one of
the PCs ($pc_l$) takes a value displaced from the optimal
one by $\sigma_{pc_l}$ and the other components vary in such
a way that the error function takes the minimum value.
As $\sigma_{pc_l}$, we took a symmetric estimate for the
error of $pc_l$, which is a diagonal element of the matrix
inverse to the Hesse matrix, (see, e.g., [45--47]), which element
was calculated as

\begin{equation}
\beta_{lm}=\sum_{k=1}^{n_{lines}} \sum_{j=1}^{N_k} \sum_{i=1}^{m_k}
\frac {1}{\sigma_{jk}^2} \frac{\partial I_{ijk}^{mod}}{\partial pc_l}
\frac{\partial I_{ijk}^{mod}}{\partial pc_m} \hspace{2mm},
\label{allalpha}
\end{equation}
where $pc_l,pc_m$ are different PCs.
The derivatives were
calculated numerically over the entire set of model maps.
After estimating the threshold value of $\chi^2_{\sigma}$,
we constructed two-dimensional projections of the error
function and its hyperplane cross-section in the plane of different
pairs of model parameters and determined the confidence regions.
In the general case, these regions
are not symmetrical relative to the optimal parameter values.
An example of using two-dimensional projections
of the error function for estimating the confidence
ranges of model parameters in the analysis of the L1287
molecular line maps is presented in Section 3.2.

\section{ESTIMATING THE PHYSICAL PARAMETERS OF THE L1287 CORE}

\subsection{Observational Manifestations of L1287}

The dark cloud L1287 is located at a distance of $0.93\pm 0.03$~kpc [48]
and is shaped as a filament of $\sim 10$~pc in length.
A dust emission map in continuum
at a wavelength of 500~$\mu$m, which was acquired using
the Herschel telescope towards L1287 (observation ID: 1342249229 [49]),
is shown in Fig. 2 (different shades).
In the central part of the cloud, there is
a high-density core, which contains the source IRAS 00338+6312 [34].
In the core, two objects of type FU Ori (RNO 1B/1C) were also
detected [51--53], as well as a cluster of IR and radio sources,
likely associated with young stellar objects of low
and intermediate mass [54, 53].
Maser lines of water [55] and methanol molecules [56] were also detected there.
Molecular line observations [34, 57, 58] revealed a bipolar outflow
in the northeastern and southwestern directions.
Based on observations in the H$^{13}$CO$^+$(1--0) line,
it was concluded [59] that the central
part of the core contains a rotating disk of radius
$\sim 7800$~AU, with the bipolar outflow oriented along the disk axis.
Based on interferometry observations,
the inner part of the core ($\la 0.1$~pc) is highly fragmented [60, 61].
In [61], a kinematic model was proposed
for the central part of the L1287 core.
In this model, gas motions towards the center from the core's
outer regions become nonisotropic near the center and
transform into the disk rotation.

\begin{figure}[t!]
\setcaptionmargin{5mm}
\onelinecaptionsfalse
	\includegraphics[width=0.85\linewidth]{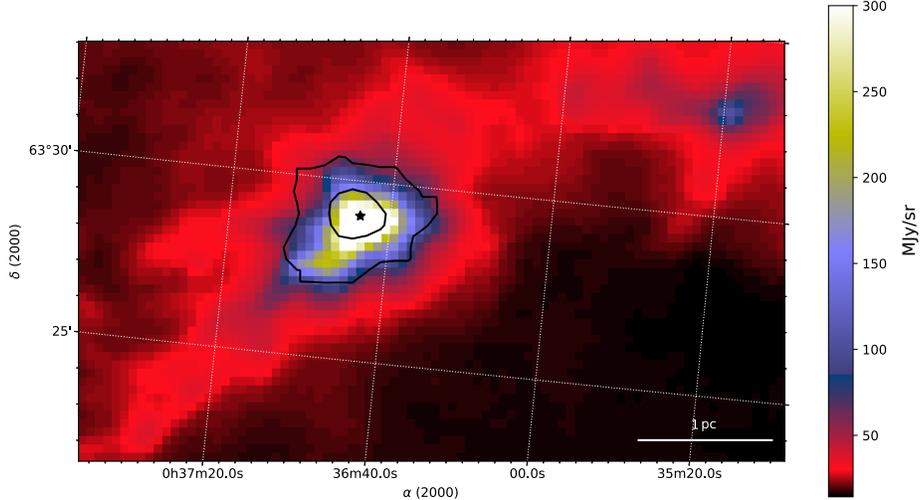}
\captionstyle{normal}
\caption{
Map of the L1287 dark cloud at wavelength 500~$\mu$m from
the Herschel telescope observational data. The integrated intensity
isolines in the HCO$^+$(1--0) line correspond to 20\% and 50\% of the maximum
(38.6 K km/s) [50]. The star symbol indicates the source IRAS 00338+6312.
}
\label{fig:map}
\end{figure}

The emission region sizes of the L1287 core
in the different molecular lines and in continuum vary
from a few tenths to one parsec [34, 62--65]; the shape
of the emission regions is roughly close to a spherically symmetric one.
The profiles of optically thick lines in the L1287 core are asymmetric
and have two peaks separated by a dip, with the amplitude of the
blue peak in most of the maps being higher than that
of the red one [25, 34, 62].

We observed the L1287 core in 2015 with the OSO-20m telescope in different
lines in the frequency range of $\sim 85-89$~GHz [50].
The angular resolution of the observations was $\sim 42''$, which corresponds
to a linear resolution of $\sim 0.19$~pc.
The integrated intensity isolines of the HCO$^+$(1--0) line, which were
superimposed onto the Herschel map, are shown in Fig. 2.
The asymmetric profiles of HCO$^+$(1--0) and HCN(1--0),
observed virtually throughout the entire region
($\sim 0.9$~pc), and the symmetric profiles of optically thin
lines, which intensity peaks are close to the dips in the
profiles of optically thick lines, are likely to be indicative
of gas contraction.

\subsection{Map Analysis of the L1287 Core
in Different Molecular Lines}

The algorithm presented in Section 2 was applied
for estimating the physical parameters of the L1287
core. To this end, we performed the fitting of the maps
in the lines of HCO$^+$(1--0), H$^{13}$CO$^+$(1--0),
HCN(1--0), and H$^{13}$CN(1--0), calculated within the 1D microturbulent
model (see Appendix), into the central part
of the observed region with an angular size of 80$''$ ($\sim 0.4$~pc).
The physical parameters (density, turbulent
and systematic velocities, and kinetic temperature)
were dependent on the distance to the center, $r$, by the law
$P=P_0/(1+(r/R_0)^{\alpha_p}$), where $R_0$ is the radius of the central layer.
The free model parameters were the values $P_0$
for the radial profiles of density and turbulent
and systematic velocities ($n_0$, $V_{turb}$, $V_{sys}$, respectively);
the power-law indices $\alpha_p$
($\alpha_n$, $\alpha_{turb}$, $\alpha_{sys}$),
the relative abundances of the molecules ($X$), independent
of the radial distance; and the outer radius ($R_{max}$) of the core.

The kinetic temperature profile was set at
$T=80$~K$/(1+(r/R_0)^{0.3})$ and kept unchanged during the calculations.
Meanwhile, the kinetic temperature
varied from 40 K in the central layer to $\la 20$~K on the
periphery, which is generally consistent with estimates
based on observational data (see, e.g., [62, 63, 65, 50]).
Although the dust temperatures for L1287 from
the Herschel data are $\sim 15-24$~K
(http://www.astro.cardiff.ac.uk/research/ViaLactea/) [66], the data of
interferometric observations suggest that in the inner
regions of the L1287 core ($\la 0.1$~pc), where the fragmentation
effects are strong, the kinetic temperature
of fragments may be as high as $\sim 80-100$~K (see [60]).
Thus, in the calculations, 40~K was taken as an average
value of kinetic temperature in the central layer, the
radius of which ($R_0$) was set at $2\times 10^{16}$~cm ($\sim 1300$~AU).

The radial velocity and the core center coordinates
were estimated from the H$^{13}$CO$^+$(1--0) line.
Then, we used a map in the HCO$^+$(1--0) line to search for the
minimum of the error function.
Using a random number generator, we formed an array of 6000 parameter
values, which were randomly and equiprobably distributed
in the following ranges of eight parameters:
$n_0$=[$10^{6.5}...10^9$]~cm$^{-3}$,
$\alpha_n$=[1.3...2.5],
$V_{turb}$=[1.4...7.5]~km/s,
$\alpha_{turb}$=[0.1...0.7],
$V_{sys}$=[--1.3...--0.2]~km/s,
$\alpha_{sys}$=[--0.2...0.4],
$X$(HCO$^+$)=[$10^{-10.5}...10^{-8}$],
$R_{max}$=[$10^{17.7}-10^{19.2}$]~cm.
Although we assumed that these ranges were certain to
include the optimal parameter values for the L1287
core, their boundaries were not rigid and could be
expanded by means of the inverse transformation from
the PC space.

For each value in the parameter array, we calculated
a map in the HCO$^+$(1--0) line and the error function.
Based on the accepted criterion, $\chi^2\le \chi_{min}^2(3\sigma_{obs})$,
we selected 246 values from the initial set.
This number was enough to construct the statistics in the PC space.
For these values, we calculated a set of PCs using a
procedure from the $scikit$-$learn$ library [67].
Using the dependence of $R$, the ratio of the sum of the
diagonal components in the PC covariance matrix to
the sum of the diagonal components in the covariance
matrix of the physical parameters, on the number of
components, we estimated the minimum number of
PCs required to represent the physical parameters (Fig. 3).
Figure 3 shows that the five PCs represent to
a sufficient extent the eight physical parameters at a level of $R=0.9$.
For the five PCs, the difference after
the inverse transformation did not exceed 5\% of the
grid step for all the parameters, suggesting no distortions
in subsequent calculations and no error accumulation.
Figure 3 also shows the contribution of each
component to the relative covariance matrix of the PCs.

\begin{figure}[t!]
\setcaptionmargin{5mm}
\onelinecaptionsfalse
\includegraphics[width=0.6\linewidth]{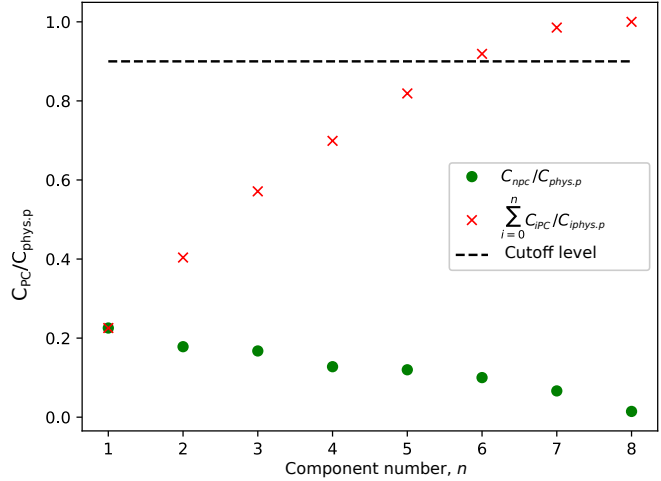}
\captionstyle{normal}
\caption{
Dependence of the ratio of the sum of the diagonal components in the PC covariance
matrix and the sum of the diagonal components in the covariance matrix
of the physical parameters on the number of components (red crosses).
The green circles show the contribution of an individual component
to the relative PC covariance matrix. The dashed horizontal line indicates
a cutoff level of 0.9.
}
\label{fig:n_comp}
\end{figure}

In the space of the five remaining PCs, we constructed
a uniform five-dimensional grid with a center
at the point of minimum of the error function; the grid
size was consistent with $6\Delta(pc_i)$, where $\Delta(pc_i)$ is a
standard deviation of the $i$th PC values, which was
calculated from the selected set of points.
The PC array was recalculated to the array of the physical
parameter values, for which we calculated the spectral
maps and estimated the error functions.
From the calculated model maps, we estimated the optimal physical
parameters from the HCO$^+$(1--0) data by the $k$NN method.
Varying by the least squares method, the relative
abundances of H$^{13}$CO$^+$, HCN, and H$^{13}$CN, we
fitted the corresponding model spectral maps into the observed ones.
In so doing, we also slightly adjusted
the parameters within the error ranges calculated from the HCO$^+$(1--0) data.
By comparing the set of model
spectral maps with the observed ones, we estimated
the global error function in the four lines by equation (3).
The resulting model spectra proved to be close to
the observed ones up to a scale of $\sim 0.8$~pc, which confirmed
the relevance of the applied model.

\begin{figure}[t!]
\setcaptionmargin{5mm}
\onelinecaptionsfalse
\includegraphics[width=\linewidth]{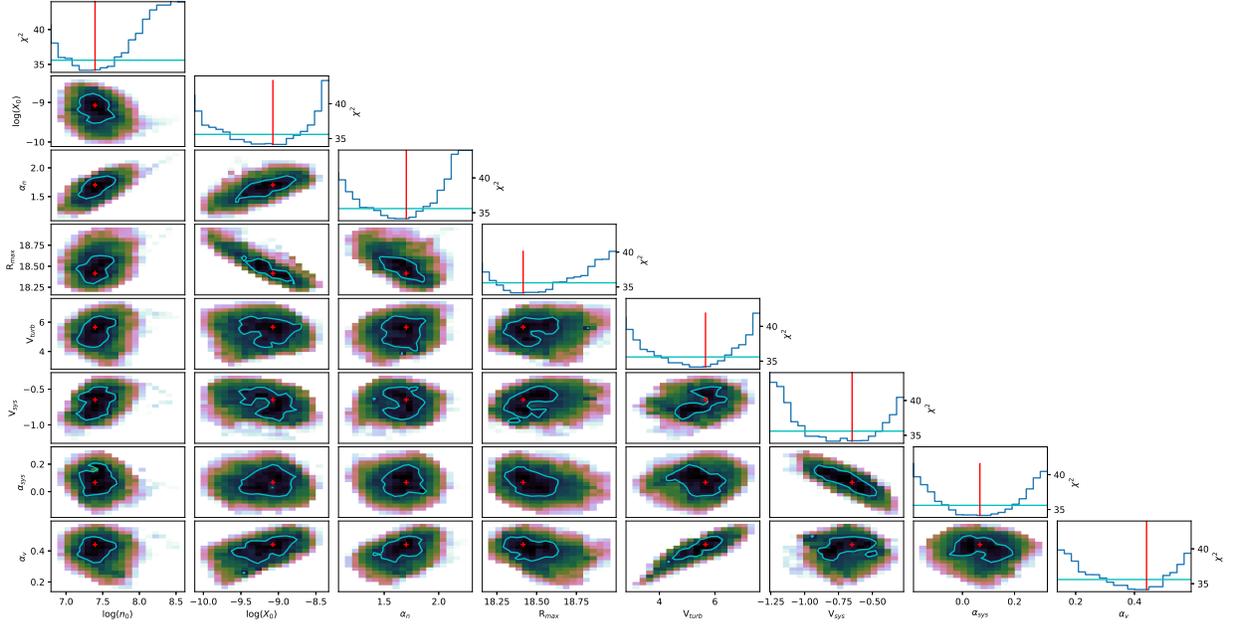}
\captionstyle{normal}
\caption{
Projections of the eight-dimensional error function $\chi^2$ onto the planes
of the different parameter pairs calculated from
the fitting of the model spectral maps in the lines HCO$^+$(1--0),
H$^{13}$CO$^+$(1--0),
HCN(1--0) and H$^{13}$CN(1--0) into the observed maps in the L1287 core.
The dependencies of the error function on the individual parameters are given
over each projection column. The red dots in the diagrams and the red vertical
lines in the upper plots indicate the global minimum of the error
function, which was obtained by $k$NN. The confidence regions for the optimal
parameter values, which were calculated from the
hyperplane $\chi_{\sigma}^2$ cross-sections of the error function, are shown
with blue contours and horizontal lines in the two-dimensional projections
and one-dimensional plots, respectively.
}
\label{fig:corner}
\end{figure}

Figure 4 presents a set of projections of the eight-dimensional
error function onto the planes of the different
parameter pairs and the error function projection dependencies
on each of the parameters.
It follows from the two-dimensional projections and the
dependencies, that the model has a global minimum, and a
confidence level can be determined for each of the parameters.
Correlations are observed between some of the parameters.
A clear correlation is observed between $R_{max}$ and the
relative abundance of HCO$^+$ ($X_0$), between $\alpha_n$ and $X_0$,
and between the turbulent and systematic velocities in the central layer
and the corresponding power-law indices of the radial
profiles of these parameters.
A weaker correlation exists between $\alpha_n$ and $n_0$ and between $R_{max}$
and $\alpha_n$.
The exact position of the minimum was estimated by the $k$NN
method from all the lines; it is marked by a red cross in
the two-dimensional projections and by red vertical
lines in the $\chi^2$ projection dependencies on individual parameters.
The confidence regions were
calculated using a cross-section of the error function by the
hyperplane $\chi_{\sigma}$.
The projections of the error function cross-sections by the $\chi_{\sigma}$
hyperplane are in fact contours in
the planes of parameter pairs; they correspond to horizontal
lines on upper plots (see Fig. 4).
The confidence regions are not symmetric with respect to the optimal values.
The distortions in the shape of the contours are due to
observational noise and the discrete filling of the parameter space.
When analyzing the dependencies of $\chi^2$
on individual parameters in broader ranges than
those shown in Fig. 4, we found additional local minima,
which values are, however, greater than the one
corresponding to the global minimum and the corresponding
parameter values considerably deviate from independent estimates.

The estimates for the physical parameters of the
L1287 core, which correspond to the global minimum
of the error function, and the uncertainties of these
estimates, which correspond to the boundaries of the
confidence regions (Fig. 4), are given in Table 1.
It should be noted that in accordance with the specified
form of the radial profiles, the physical parameter values
in the central layer are twice as low as the corresponding
values of $n_0$, $V_{turb}$, and $V_{sys}$.

\begin{table}[p]
\setcaptionmargin{0mm}
\onelinecaptionsfalse

\centering
\caption{Resulting values of the physical parameters}
\vskip 2mm
\begin{tabular}{ll}
\noalign{\hrule}\noalign{\smallskip}
Parameter & Value \\
\noalign{\hrule}\noalign{\smallskip}
$n_0$(cm$^{-3}$), 10$^7$  & 2.6$_{-1.3}^{+1.7}$   \\
$\alpha_n$            & 1.7$_{-0.3}^{+0.1}$   \\
V$_{turb}$ (km/s)          & 5.6$_{-1.4}^{+0.7}$   \\
$\alpha_{turb}$        &0.44$_{-0.13}^{0.05}$   \\
V$_{sys}$ (km/s)            &--0.66$_{-0.24}^{+0.21}$\\
$\alpha_{sys}$        &0.1$_{-0.13}^{+0.08}$ \\
R$_{max}$(pc)              &0.8$_{-0.25}^{+0.2}$    \\
X(HCO$^+$), 10$^{-9}$         &1.0$_{-0.4}^{+0.5}$ \\
X(H$^{13}$CO$^+$), 10$^{-11}$ &3.7$_{-2.0}^{+2.4}$ \\
X(HCN), 10$^{-9}$             &2.5$_{-1.1}^{+1.4}$ \\
X(H$^{13}$CN), 10$^{-11}$     &8.5$_{-4.8}^{+5.3}$ \\
\noalign{\smallskip}\hline\noalign{\smallskip}
\end{tabular}
\label{table:physpar}
\end{table}

\section{RESULTS AND DISCUSSION}

Figures 5 and 6 show the spectral maps for the central
part of the L1287 core ($\sim 0.4$~pc) in the HCO$^+$(1--0), HCN(1--0),
H$^{13}$CO$^+$(1--0), and H$^{13}$CN(1--0) lines with the fitted model
spectra corresponding to the global minimum of the error function.
The asymmetry and dip in the observed profiles of the
optically thick lines of HCO$^+$(1--0) and HCN(1--0)
are well reproduced by the model.
In the central and southwestern parts of the analyzed region, the spectra
of the optically thick lines exhibit high-velocity gas
emission, which was ignored in the model calculations.
The slight discrepancy between the model and observed spectra at the edges
of the observed region may be due to a difference from
spherical symmetry. The diameter (1.6 pc) of the
model cloud exceeds the observed sizes of the
emission regions in the different molecular lines, the
dense gas indicators, HCO$^+$(1--0), HCN(1--0), and
NH$_3$(1,1) ($\sim 0.3-0.5$~pc) [62, 63, 50], since it comprises
the low-density outer layers, which cause the dip in the profiles
of the optically thick lines as they absorb the emission from the
central layers.

\begin{figure}[t!]
\setcaptionmargin{5mm}
\onelinecaptionsfalse
    \includegraphics[width=\linewidth]{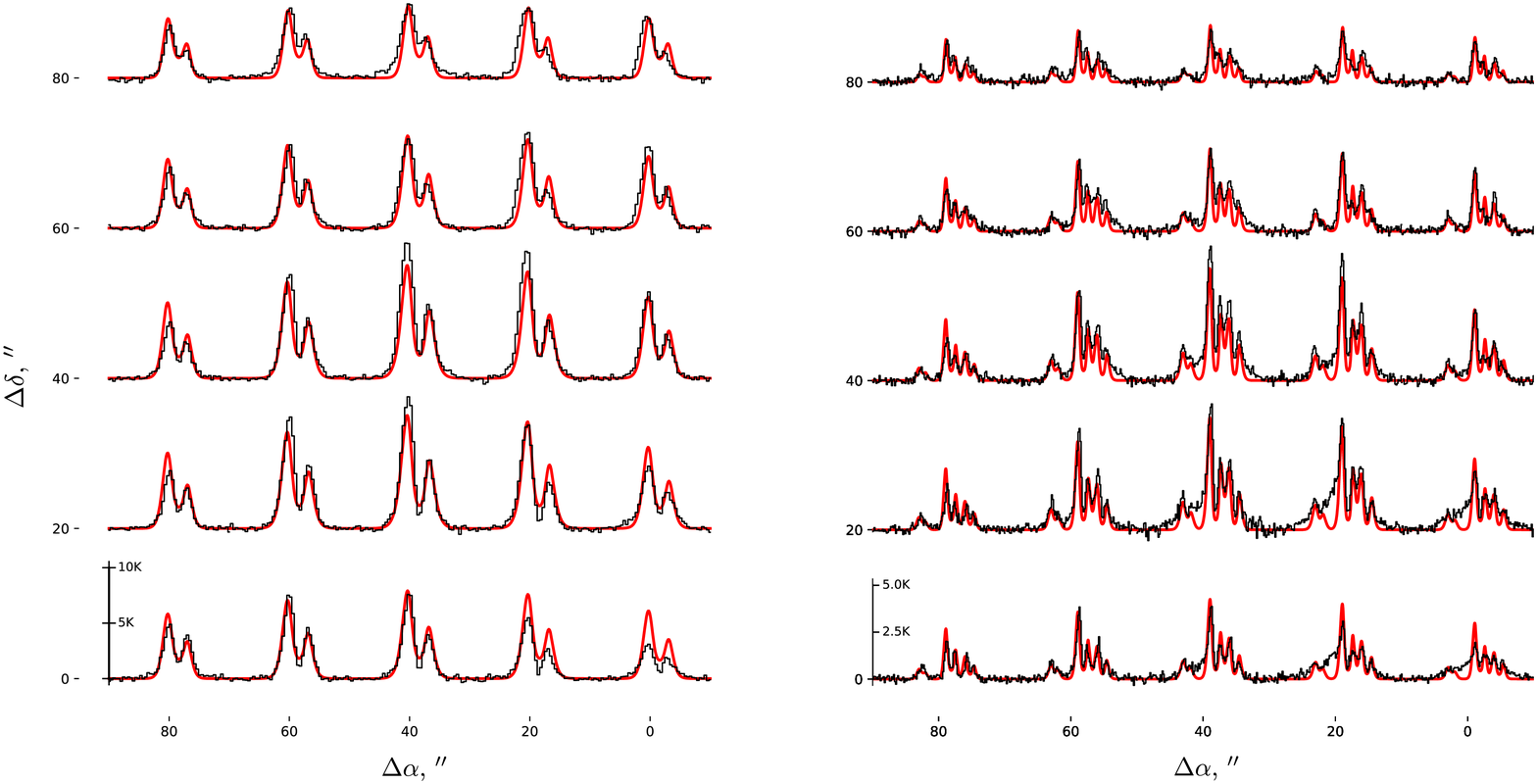}
\captionstyle{normal}
\caption{
Results of fitting the model spectra of HCO$^+$(1--0) (left) and HCN(1--0)
(right) (smooth red curves) into the observed
ones (histograms, black lines) in the central part of the L1287 core.
The horizontal axis plots the velocities in the range from --33
to --5 km/s.
}
\label{fig:specmap}
\end{figure}

\begin{figure}[t!]
\setcaptionmargin{5mm}
\onelinecaptionsfalse
    \includegraphics[width=\linewidth]{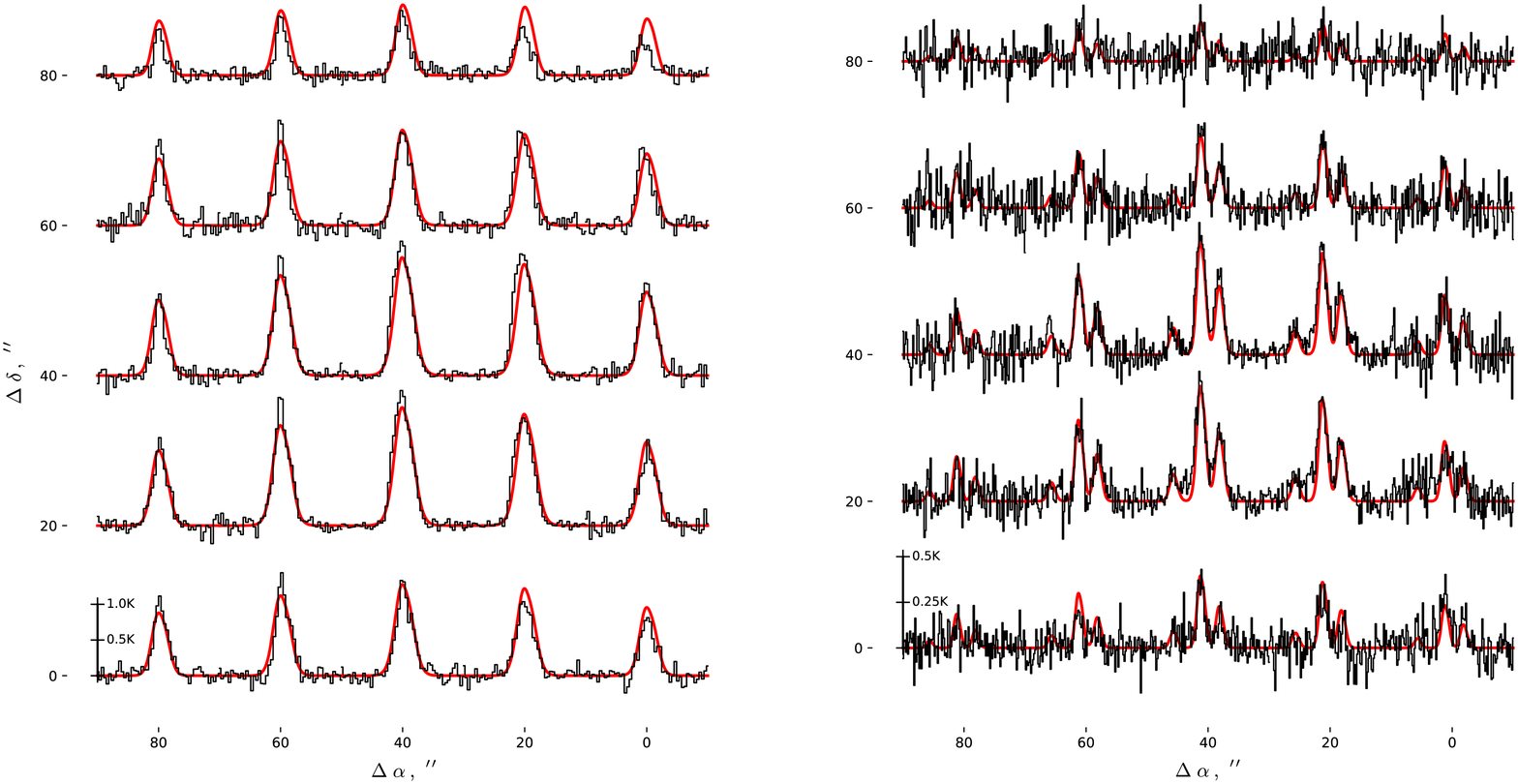}
\captionstyle{normal}
\caption{
Results of fitting the model spectra of H$^{13}$CO$^+$(1--0) (left)
and H$^{13}$CN(1--0) (right) (smooth red curves) into the observed
ones (histograms, black lines) in the central part of the L1287 core.
The horizontal axis plots the velocities in the range from --28.5
to --7 km/s.
}
\label{fig:specmap1}
\end{figure}

The calculated physical parameters of the core are
consistent, considering the errors (see Table 1), with
estimates obtained from the data of independent observations.
Thus, the model column density of
molecular hydrogen for a region of radius $\sim 20''$ agrees
with the value calculated from the data of dust observations
with the Herschel telescope [66]
($4.6_{-2.3}^{+3.0}$\,$10^{23}$~cm$^{-2}$ and $(1.8\pm 1.2)\,10^{23}$~cm$^{-2}$,
respectively).
The core mass calculated from the model for a region
of radius $\sim 0.6$~pc is $\sim 1200$~M$_0$; considering the error
($\ga 50$\%), associated primarily with the error of $n_0$, this
mass is consistent with the value of 810~M$_0$, obtained
from the observations of dust for a region of similar radius [65].
Neither does the power-law index of the radial density profile $1.7_{-0.3}^{+0.1}$
contradict the value of $1.25\pm 0.2$, obtained from the observations
of dust in continuum [65].
Both of these estimates lie in the
value range for the power-law index of the density profile
obtained for samples of dense cores associated
with regions of massive star and star
cluster formation (see, e.g., [65, 68, 69]) but are lower than 2,
the value predicted by the isothermal sphere [8]
and global collapse models [13].

The model abundance ratios of the main and rarer
isotopes ([HCO$^+$]/[H$^{13}$CO$^+$] and [HCN]/[H$^{13}$CN])
are lower by a factor of $\sim 2$ than the isotope ratio
[$^{12}$C]/[$^{13}$C]$\sim 58$, calculated from the heliocentric
dependence of this ratio [70] for $R_G\sim 9$~kpc (L1287).
However, the uncertainties of the model abundance
ratios are rather high ($\ga 80$\%) to make further
conclusions from this discrepancy.
To verify the results obtained, they should be compared with the chemical
modeling results.

The turbulent velocity falls rather sharply with the
distance from the center (from 2.8 km/s in the central
layer to $\sim 0.6$~km/s in the outer layer), which is necessary
for reproducing the shape of the dip in the
HCO$^+$(1--0) and HCN(1--0) spectra (solid curve $1$ in Fig. 7, right panel).
The contraction velocity decreases
weakly in absolute terms with the distance from the
center ($\sim 0.33$~km/s in the central layer and $\sim 0.25$~km/s
in the outer layer) (dashed curve $1$ in Fig. 7, right panel).
Its average value across the model cloud is $0.26\pm 0.09$~km/s, which
does not contradict the value of $\sim 0.22$~km/s, calculated
from the HCO$^+$(1--0) line parameters for the point (60$''$,40$''$)
by the formula of the two-layer model [20] (the value given in [50] is
underestimated).

\begin{figure}[t!]
\setcaptionmargin{5mm}
\onelinecaptionsfalse

\begin{minipage}[b]{0.49\textwidth}
	\includegraphics[width=\linewidth]{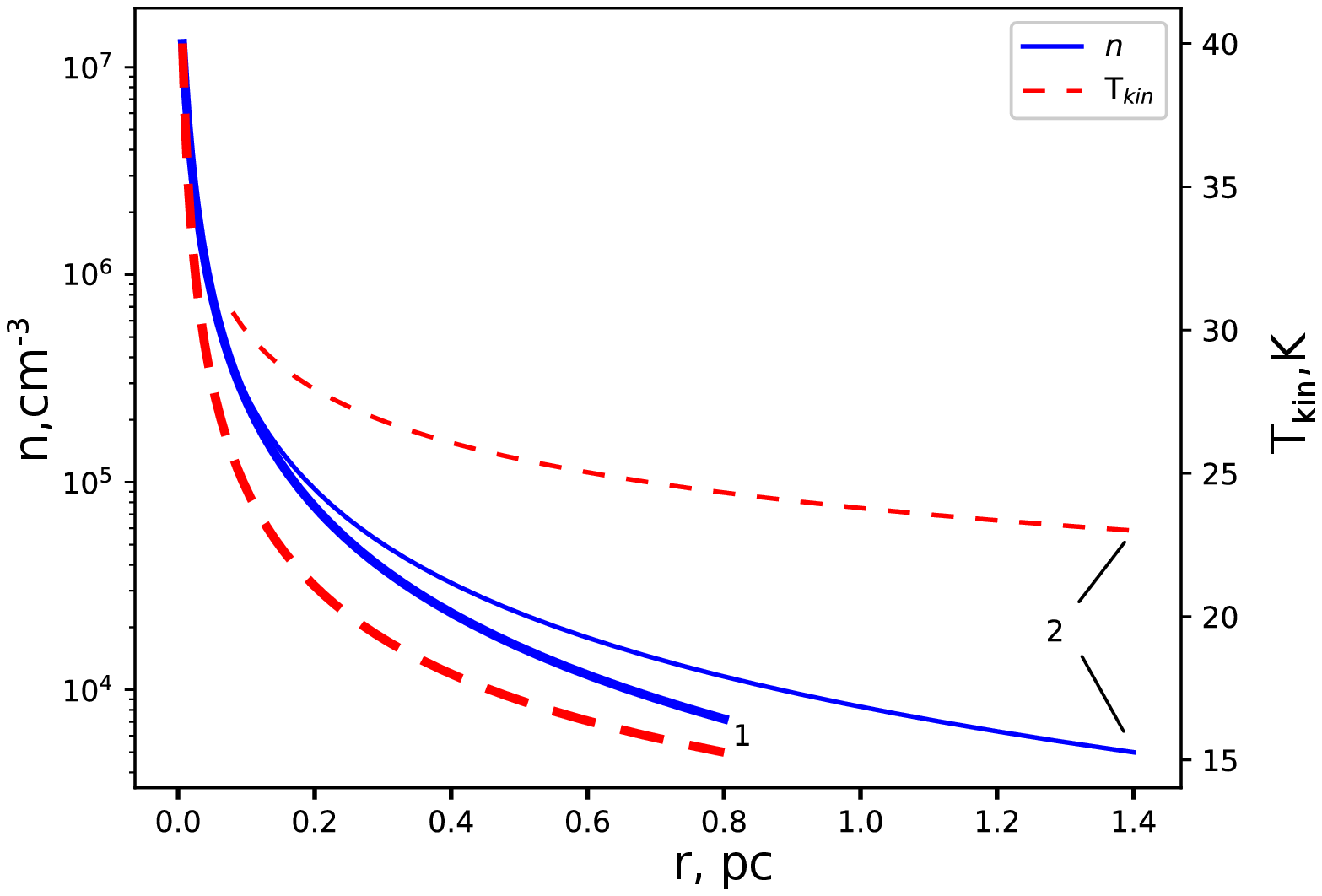}
\end{minipage}
\begin{minipage}[b]{0.49\textwidth}
	\includegraphics[width=\linewidth]{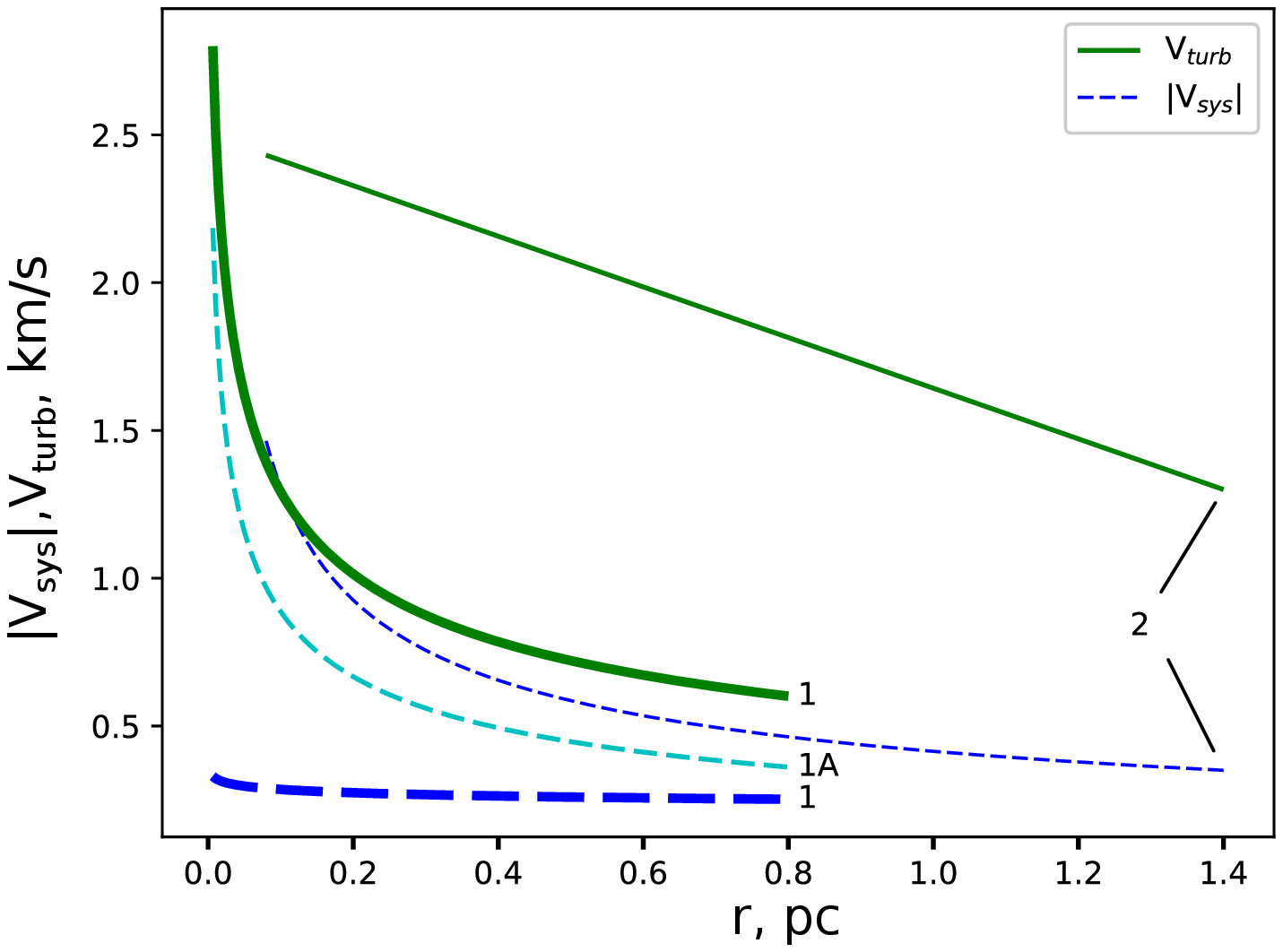}
\end{minipage}

\captionstyle{normal}
\caption{
Model radial profiles of density and kinetic temperature (left panel) and
contraction and turbulent velocities (right panel).
Profiles were obtained in this work ($1$) and those from [25] ($2$).
The cloud radius in the latter model was set to 1.4~pc,
consistent with a distance of 0.93~kpc to L1287.
The $1A$ mark in the right panel shows the contraction velocity profile
obtained in our model with a fixed value of the power index, 0.5.
}
\label{fig:profiles}
\end{figure}

The power-law index of the radial profile for contraction
velocity obtained in the model calculations
considering errors proved to be lower than 0.5 for the case
of gas collapse onto the protostar in free fall [8, 10, 13].
In [25], the observational data for the core 0038 + 6312
(L1287) in the HCO$^+$(1--0), CS(2--1), and CS(5--4)
lines were compared with the calculated results for the
model with the density and contraction velocity profiles $\propto r^{-3/2}$
and $\propto r^{-1/2}$, respectively.
Although the intensities and widths of the model spectra proved to
match the observed ones in the
direction of individual positions quite well, considering the sensitivity
and spectral resolution of the data, the dip in the HCO$^+$(1--0) line profiles
was not reproduced (see [25]).
This is due to the difference of the radial profiles for velocity, which were
assumed in the model [25], from the profiles obtained in our calculations
(Fig. 7, right panel).
The left panel in Fig. 7
shows the radial profiles of density and kinetic temperature for
our model and for the model [25].

As shown in the model of the global hierarchical
collapse (see, e.g., [13, 16]), if the core is globally
out-of-equilibrium, it experiences contraction with a constant
velocity and this contraction continues in the
outer layers even after the protostar formation.
For comparison, we fitted the model maps into the
observed ones for the case where the power-law index
of the radial profile of systematic velocity was fixed at 0.5.
The corresponding velocity profile is marked $1A$ in Fig. 7 (right panel).
Figure 8 shows the observed
HCO$^+$(1--0) and HCN(1--0) lines for the point (60$''$,40$''$) near the core
center and the model spectra for the power-law index of systematic velocity
of 0.1, which corresponds to the global minimum of the error function,
and for the case when the index is 0.5, respectively.
Upon comparison of the spectra, the model
with the index 0.1 more accurately reproduces the
intensities and widths of the asymmetric HCO$^+$(1--0)
profiles and, specifically, the profiles of three
hyperfine HCN(1--0) components than the model with the index 0.5.
A similar conclusion is true for the other points.
Although in the southwestern part, the high-velocity
emission associated with bipolar outflow more
strongly distorts the spectra shape (Fig. 5), which
makes it more difficult to compare the models.
The fact that the value obtained in the
model with the power-law index of the velocity profile as a free parameter
turned out to be lower than 0.5 may suggest a likelihood
of nonuniform gas contraction in the core -- with constant
velocity in the extended envelope and with the $\propto r^{-1/2}$
profile in the region near the center.
The use of the model with a single power-law index gives a
weighted average for the entire core in this case.

\begin{figure}[t!]
\setcaptionmargin{5mm}
\onelinecaptionsfalse

\begin{minipage}[b]{0.45\textwidth}
	\includegraphics[width=\linewidth]{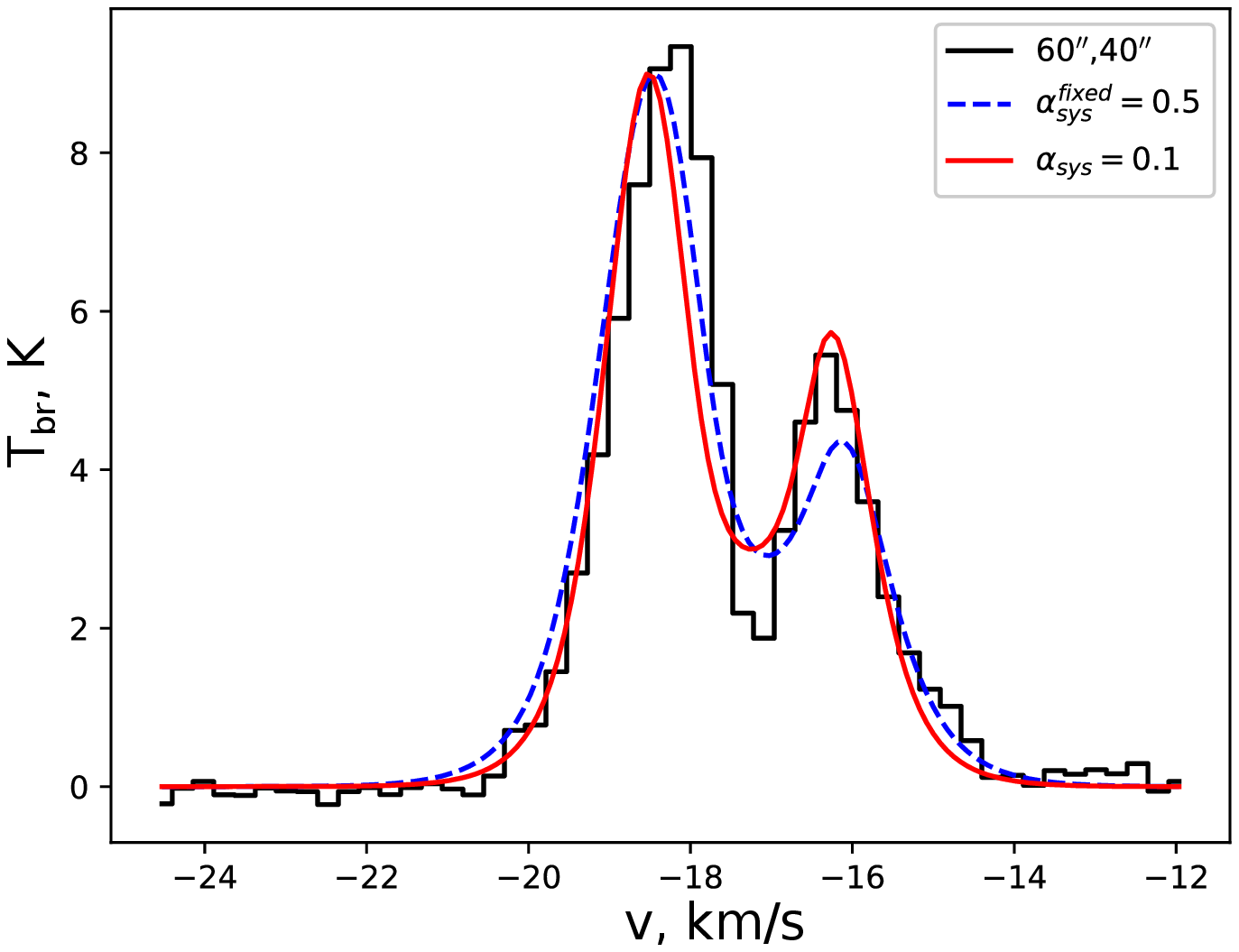}
\end{minipage}
\begin{minipage}[b]{0.45\textwidth}
	\includegraphics[width=\linewidth]{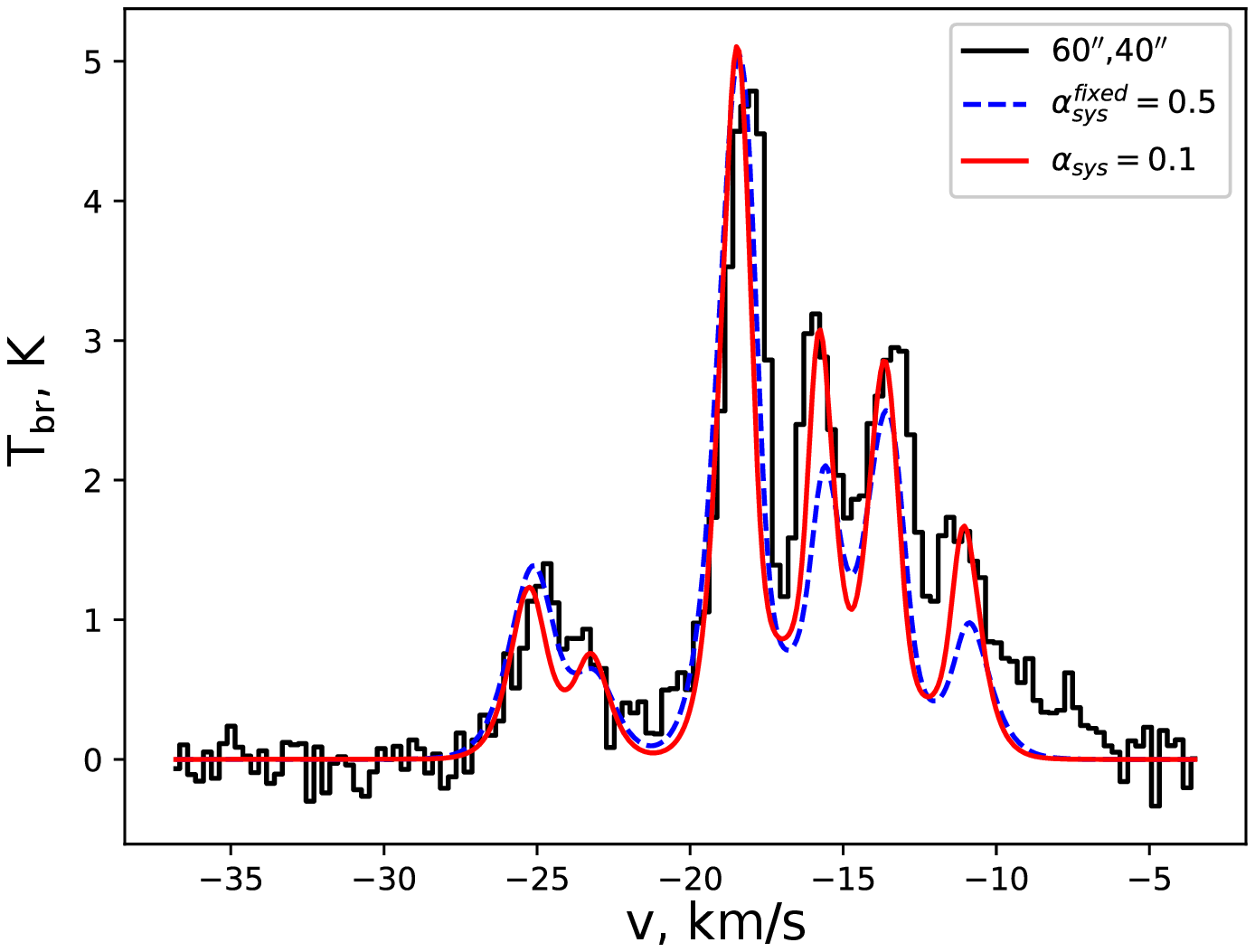}
\end{minipage}

\captionstyle{normal}
\caption{
Observed and model profiles of the lines HCO$^+$(1--0) (left) and HCN(1--0)
(right) towards the (60$''$,40$''$) position for models
with different values of the power-law index in the radial profile
of contraction velocity.
}
\label{fig:vsys}
\end{figure}

Although we used a rather simplified 1D model with uniform
power-law indices for the radial profiles of the
physical parameters, it allows using the developed
algorithm for fitting the model spectral maps
into the observed ones with PCA and $k$NN  to reproduce accurately
the observed HCO$^+$(1--0), HCN(1--0), H$^{13}$CO$^+$(1--0),
and H$^{13}$CN(1--0) line maps and estimate the radial profiles of
the parameters in outer regions of the L1287 core ($r\ga 0.1$~pc).
Some of the differences between the
model and observed spectra can be eliminated, apparently,
within more complex models with composite
radial profiles of the parameters and also within 2D or 3D
models considering the possible spatial inhomogeneity
of the velocity field, rotation, and high-velocity outflows.
To reduce errors in the calculated parameters
and to confirm the conclusion about the possible
global contraction of the L1287 core, further observations
are required in molecular lines of different optical
depths with better spatial and spectral resolution.

A distinctive feature of the developed algorithm is
its lack of ties to a specific model and its capability of
simultaneous analysis of spectral maps with different
spatial resolutions and sizes, as well as maps in continuum.
The main constraint is only the computational
time required to construct the necessary statistics.

\section{CONCLUSIONS}

The results of this work can be summarized as follows:

(1) An algorithm was developed for finding the
global minimum of the multidimensional error function
and calculating the optimal parameter values
when fitting model spectral maps into observed ones.
The algorithm is based on applying
principal component analysis to a given range of parameters,
as a result of which a reduction is achieved in the model's dimensionality
and in the coupling degree between the
parameters. Thus, the region of the minimum is determined.
The optimal parameter values are calculated
using the $k$ nearest neighbors method.
Confidence regions for the
optimal parameter values are determined using a cross-section
of the error function by a hyperplane calculated in
the PC space and its projections onto the various pairs
of the parameters. The algorithm is not tied to a specific
model.

(2) The algorithm was used to perform the fitting of
the model maps in the HCO$^+$(1--0), H$^{13}$CO$^+$(1--0), HCN(1--0),
and H$^{13}$CN(1--0) lines into the observed
maps of the protostellar core of L1287, in which the
formation of a young stellar cluster is underway, and
the asymmetry of the profiles of optically thick
lines indicates contraction.
The maps were calculated
within a spherically symmetric 1D model in which the
physical parameters (density and turbulent and systematic
velocities) were functions of the distance from
the center.
Optimal values were calculated for the
model parameters, and their errors were determined.
It was found that density in the L1287 core decreases
with the distance from the center as $\propto r^{-1.7}$ while turbulent
and contraction velocities decrease as $\propto r^{-0.4}$
and $\propto r^{-0.1}$, respectively.
The absolute value of the power-law
index for the radial profile of contraction velocity,
considering the probable error, is less than 0.5, a value
expected in the case of gas collapse onto the protostar
in free fall. This result may indicate global contraction
in the L1287 core, which was predicted in several theoretical
works.

\vskip 0.5 cm

\begin{center}
{\bf APPENDIX. MODEL DESCRIPTION}
\end{center}

Excitation of rotational levels of the HCO$^+$ and HCN
molecules and their isotopes and the profiles of the
(1--0) transitions were calculated within a spherically
symmetric microturbulent model.
The model cloud is a set of concentric layers in which a certain
physical parameter (density, kinetic temperature,
turbulent and systematic velocity) was set constant,
changing from one layer to another by the relationship
$P=P_0/(1+(r/R_0)^{\alpha_p}$), where $r$ is the distance from the
center and $R_0$ is the radius of the central layer.
This functional dependence, which is a simplified form of
the Plummer function, is used quite often as a model
density profile (see, e.g., [22]) to avoid singularity at the center.
In our model, this form of the dependence was used for all the parameters.
The values of $P_0$ and $\alpha_p$ for each parameter were varied
while fitting the model profiles into the observed ones.
The kinetic temperature profile was taken as
$T=80$~K$/(1+(r/R_0)^{0.3})$
and remained unchanged during calculations.
It should be noted that kinetic temperature affects the intensities of
the calculated HCO$^+$(1--0) and HCN(1--0) lines to a
lesser degree than density and concentration.
Turbulent velocity was a parameter that gives an additional
contribution - aside from the thermal one - to the
local width of the lines.
The relative molecular abundance was independent of radial distance.
When calculating the excitation of HCN and H$^{13}$CN, the
hyperfine structure of the rotational spectrum and the
overlapping of closely located hyperfine components
[40, 71] was taken into account.
The description of the
model and calculation techniques for radiation transfer
in the case of HCN is given in the Appendix to [40].
In our version of this model, the layer width
increases by the power law with the distance from the
center, and the radial profile of systematic velocity,
which gives a Doppler shift to the local profile of the
line, is taken into account. The calculations were conducted
for 14 layers. The calculations used collisional
probabilities of HCO$^+$--H$_2$ [72] and HCN--H$_2$
taking into account hyperfine structure [73].

Excitation of rotational levels of a certain molecule
was calculated by an iterative method, sequentially for
one point in each layer, the radial distance of which is
equal to the geometric mean of the inner and outer
radii of the layer. To this end, a system of population
balance equations was solved, while the populations in
other layers were considered unchanged. After reaching
the outer layer, the populations in each layer were
compared with the values obtained in the previous
iteration, and the process was repeated [40]. To
increase the accuracy of calculating the radiation
transfer in a moving medium, each layer was additionally
divided into ten sublayers, with different systematic
velocities. A test comparison of the calculated
results for this model with the calculated results in [74]
for a molecule with two energy levels showed that the
calculated populations differ by no more than 0.4\% in
the case of line optical depth of $\la 60$.

The model code, written in Fortran, was controlled
by means of a module written in Python. Model spectra
were calculated for the different impact parameters.
Using the astropy.convolve\_fft procedure [75],
the resulting maps were convoluted channel by channel
with a two-dimensional Gaussian of width 40$''$
(the width of the main beam of the OSO-20m radio
telescope at a frequency of $\sim 90$~GHz). The model
spectra were fitted into the observed ones using a
PCA- and $k$NN-based algorithm (Section 2), written
in Python.

\begin{acknowledgments}

The authors thank the reviewer Ya.N. Pavlyuchenkov
for his valuable remarks and additions.

This work was supported by the Russian Science Foundation,
project no. 17-12-01256 (analysis of the results), and
the Russian Foundation for Basic Research, project no. 18-
02-00660-a) (program development and model calculations).

\end{acknowledgments}

\begin{center}
{\bf REFERENCES}
\end{center}

1. J. C. Tan, M. Beltr\'an, P. Caselli, F. Fontani, A. Fuente,
M. R. Krunholz, C. F. McKee, and A. Stolte, 
in $Protostars~and~Planets~VI$, Ed. by H. Beuther, R. S. Klessen,
C. P. Dullemond, and Th. Henning (Univ. of Arizona
Press, Tucson, 2014), p. 149.

2. F. Motte, S. Bontemps, and F. Louvet, Ann. Rev. Astron.
Astrophys. 56, 41 (2018).

3. Ph. Andr\'e, A. Men'shchikov, S. Bontemps, V. K\"onyves,
et al., Astron. Astrophys. 518, L102 (2010).

4. Ph. Andr\'e, J. Di Francesco, D. Ward-Thompson,
S.-I. Inutsuka, R. E. Pudritz, and J. E. Pineda, 
in $Protostars~and~Planets~VI$, Ed. by H. Beuther, R. S. Klessen,
C. P. Dullemond, and Th. Henning 
(Univ. of Arizona Press, Tucson, 2014), p. 27.

5. F. Nakamura, K. Sugitani, T. Tanaka, H. Nishitani,
et al., Astrophys. J. Lett. 791, L23 (2014).

6. Y. Contreras, G. Garay, J. M. Rathborne, and P. Sanhueza,
Mon. Not. R. Astron. Soc. 456, 2041 (2016).

7. L. K. Dewangan, L. E. Pirogov, O. L. Ryabukhina,
D. K. Ojha, and I. Zinchenko, Astrophys. J. 877, 1
(2019).

8. F. H. Shu, Astrophys. J. 214, 488 (1977).

9. F. H. Shu, F. C. Adams, and S. Lizano, Ann. Rev.
Astron. Astrophys. 25, 23 (1987).

10. C. F. McKee and J. C. Tan, Astrophys. J. 585, 850
(2003).

11. Y. Zhang and J. C. Tan, Astrophys. J. 853, 18 (2018).

12. D. E. McLaughlin and R. E. Pudritz, Astrophys. J. 476,
750 (1997).

13. E. V\'azquez-Semadeni, A. Palau, J. Ballesteros-Paredes,
G. C. G\'omez, and M. Zamora-Avil\'es, 
Mon. Not. R. Astron. Soc. 490, 3061 (2019).

14. R. B. Larson, Mon. Not. R. Astron. Soc. 145, 271 (1969).

15. M. V. Penston, Mon. Not. R. Astron. Soc. 144, 425 (1969).

16. R. Naranjo-Romero, E. V\'azquez-Semadeni, and
R. M. Loughnane, Astrophys. J. 814, 48 (2015).

17. D. Mardones, P. C. Myers, M. Tafalla, D. J. Wilner,
R. Bachiller, and G. Garay, Astrophys. J. 489, 719 (1997).

18. N. J. Evans II, Ann. Rev. Astron. Astrophys. 37, 311 (1999).

19. Ya. Pavlyuchenkov, D. Wiebe, B. Shustov, Th. Henning,
R. Launhardt, and D. Semenov, Astrophys. J. 689, 335 (2008).

20. P. C. Myers, D. Mardones, M. Tafalla, J. P. Williams,
and D. J. Wilner, Astrophys. J. 465, L133 (1996).

21. C. W. Lee, P. C. Myers, and M. Tafalla, Astrophys. J.
Suppl. 136, 703 (2001).

22. C. H. De Vries and P. C. Myers, Astrophys. J. 620, 800 (2005).

23. J. L. Campbell, R. K. Friesen, P. G. Martin, P. Caselli,
J. Kauffmann, and J. E. Pineda, Astrophys. J. 819, 143 (2016).

24. S. Zhou, N. J. Evans, II, C. K\"ompe, and C. M. Wamsley,
Astrophys. J. 404, 232 (1993).

25. R. N. F. Walker and M. R. W. Masheder, Mon. Not. R.
Astron. Soc. 285, 862 (1997).

26. G. Narayanan, G. Moriarty-Schieven, C. K. Walker,
and H. M. Butner, Astrophys. J. 565, 319 (2002).

27. Ya. N. Pavlyuchenkov and B. M. Shustov, Astron. Rep. 48, 315 (2004).

28. L. Pagani, I. Ristorcelli, N. Boudet, M. Giard, A. Abergel,
and J.-P. Bernard, Astron. Astrophys. 512, A3 (2010).

29. G. Garay, D. Mardones, Y. Contreras, J. E. Pineda,
E. Servajean, and A. E. Guzm\'an, Astrophys. J. 799, 75 (2015).

30. K. R. Opara and J. Arabas, Swarm Evol. Comput. 44, 546 (2019).

31. B. Sch\"olkopf, A. Smola, and K.-R. M\"uller, in Artificial
Neural Networks, Proceedings of the ICANN 97 (1997), p. 583.

32. M. H. Heyer and P. Schloerb, Astrophys. J. 475, 173 (1997).

33. N. S. Altman, Am. Statist. 46, 175 (1992).

34. J. Yang, T. Umemoto, T. Iwata, and Y. Fukui, Astrophys. J. 373, 137 (1991).

35. P. D. Klaassen, L. Testi, and H. Beuther, Astron.
Astrophys. 538, A140 (2012).

36. F. Wyrowski, R. G\"usten, K. M. Menten, H. Wiesemeyer,
et al., Astron. Astrophys. 585, A149 (2016).

37. Y.-X. He, J.-J. Zhou, J. Esimbek, W.-G. Ji, et al., Mon.
Not. R. Astron. Soc. 461, 2288 (2016).

38. H. Yoo, K.-T. Kim, J. Cho, M. Choi, J. Wu, N. J. Evans
II, and L. M. Ziurys, Astrophys. J. Suppl. 235, 31 (2018).

39. N. Cunningham, S. L. Lumsden, T. J. T. Moore,
L. T. Maud, and I. Mendigutia, Mon. Not. R. Astron.
Soc. 477, 2455 (2018).

40. B. E. Turner, L. Pirogov, and Y. C. Minh, Astrophys. J. 483, 235 (1997).

41. L. Pirogov, Astron. Astrophys. 348, 600 (1999).

42. R. Cangelosi and F. Goriely, Biol. Direct 2, 2 (2007).

43. I. T. Jolliffe, Principal Component Analysis, Springer Series
in Statistics (Springer, New York, 2002).

44. P. M. Zemlyanukha, I. I. Zinchenko, S. V. Salii,
O. L. Ryabukhina, and Sh.-Yu. Liu, Astron. Rep. 62, 326 (2018).

45. G. T. Smirnov and A. P. Tsivilev, Sov. Astron. 26, 616 (1982).

46. W. H. Press, S. A. Teukolsky, W. T. Vetterling, and
B. P. Flannery, Numerical Recipes in Fortran 77 (Cambridge
Univ. Press, Cambridge, 1992).

47. S. Brandt, $Data~Analysis:~Statistical~and~Computational~
Methods~for~Scientists\\
~and~Engineers$ (Springer Nature,
Switzerland, 2014).

48. K. L. J. Rygl, A. Brunthaler, M. J. Reid, K. M. Menten,
H. J. van Langevelde, and Y. Xu, Astron. Astrophys. 511, A2 (2010).

49. S. Molinari, B. Swinyard, J. Bally, M. Barlow, et al.,
Astron. Astrophys. 518, L100 (2010).

50. L. E. Pirogov, V. M. Shul'ga, I. I. Zinchenko, P. M. Zemlyanukha,
A. H. Patoka, and M. Thomasson, Astron. Rep. 60, 904 (2016).

51. H. J. Staude and T. Neckel, Astron. Astrophys. 244, L13 (1991).

52. S. McMuldroch, G. A. Blake, and A. I. Sargent,
Astron. J. 110, 354 (1995).

53. S. P. Quanz, Th. Henning, J. Bouwman, H. Linz, and
F. Lahuis, Astrophys. J. 658, 487 (2007).

54. G. Anglada, L. F. Rodr\'iguez, J. M. Girart, R. Estalella,
and J. M. Torrelles, Astrophys. J. 420, L91 (1994).

55. D. Fiebig, Astron. Astrophys. 298, 207 (1995).

56. C.-G. Gan, X. Chen, Z.-Q. Shen, Y. Xu, and B.-G. Ju,
Astrophys. J. 763, 2 (2013).

57. R. L. Snell, R. L. Dickman, and Y.-L. Huang, Astrophys.
J. 352, 139 (1990).

58. Y. Xu, Z.-Q. Shen, J. Yang, X. W. Zheng, et al., Astron.
J. 132, 20 (2006).

59. T. Umemoto, M. Saito, J. Yang, and N. Hirano, in $Star~
Formation$ 1999,Ed. by T. Nakamoto (Nobeyama Radio
Observatory, 1999), p. 227.

60. O. Feh\'er, A. K\'osp\'al, P. \'Abrah\'am, M. R. Hogerheijde,
and C. Brinch, Astron. Astrophys. 607, A39 (2017).

61. C. Ju\'arez, H. B. Liu, J. M. Girart, A. Palau,
G. Busquet, R. Galv\'an-Madrid, N. Hirano, and
Y. Lin, Astron. Astrophys. 621, A140 (2019).

62. R. Estalella, R. Mauersberger, J. M. Torrelles, G. Anglada,
J. F. G\'omez, R. L\'opez, and D. Muders, Astrophys. J. 419, 698 (1993).

63. I. Sep\'ulveda, G. Anglada, R. Estalella, R. L\'opez,
J. M. Girart, and J. Yang, Astron. Astrophys. 527, A41 (2011).

64. G. Sandell and D. A. Weintraub, Astrophys. J. Suppl. 134, 115 (2001).

65. K. E. Mueller, Y. L. Shirley, N. J. Evans II, and
H. R. Jacobson, Astrophys. J. Suppl. 143, 469 (2002).

66. K. A. Marsh, A. P. Whitworth, and O. Lomax, Mon.
Not. R. Astron. Soc. 454, 4282 (2015).

67. F. Pedregosa, G. Varoquaux, A. Gramfort, and V. Michel,
J. Mach. Learn. Res. 12, 2825 (2011).

68. S. J. Williams, G. A. Fuller, and T. K. Sridharan,
Astron. Astrophys. 434, 257 (2005).

69. L. E. Pirogov, Astron. Rep. 53, 1127 (2009).

70. Y. T. Yan, J. S. Zhang, C. Henkel, T. Mufakharov,
et al., Astrophys. J. 877, 154 (2019).

71. S. Guilloteau and A. Baudry, Astron. Astrophys. 97, 213 (1981).

72. D. R. Flower, Mon. Not. R. Astron. Soc. 305, 651 (1999).

73. D. Ben Abdallah, F. Najar, N. Jaidane, F. Dumouchel,
and F. Lique, Mon. Not. R. Astron. Soc. 419, 2441
(2012),

74. G.-J. van Zadelhoff, C. P. Dullemond, F. F. S. van der
Tak, J. A. Yates, et al., Astron. Astrophys. 395, 373
(2002).

75. The Astropy Collaboration, T. P. Robitaille, E. J. Tollerud, P. Greenfield, 
M. Droettboom, et al., Astron. Astrophys. 558, A33 (2013).

\vskip 0.5 cm
\hskip 11 cm
$Translated~by~A.~Kobkova$

\end{document}